\documentclass[11pt]{article}
\usepackage{amsfonts,amsmath,amssymb,mathrsfs}
\usepackage{color}
\usepackage{epsfig,graphicx}
\usepackage{hyperref}

\newtheorem{theorem}{Theorem}

\newtheorem{remark}{Remark}
\newtheorem{problem}{Problem}
\newtheorem{example}{Example}

\title{Coherent Feedback Control of Linear Quantum Optical Systems via Squeezing and Phase Shift\thanks{%
This work was partially supported by the National Natural Science Foundation
of China under Grant 60804015,  RGC PolyU 5203/10E, AFOSR Grant FA2386-09-1-4089 AOARD 094089.}}

\author{Guofeng~Zhang\thanks{G. Zhang is with the Department of Applied Mathematics, the Hong Kong
Polytechnic University, Hong Kong, China, and he was with the College of Engineering and Computer Science,
the Australian National University, Canberra, ACT 0200, Australia (e-mail: Guofeng.Zhang@polyu.edu.hk).} \and Heung Wing Joseph Lee\thanks{H.W.J. Lee is with the Department of Applied Mathematics, the Hong Kong
Polytechnic University, Hong Kong, China (email: majlee@polyu.edu.hk).} \and
Bo Huang\thanks{B. Huang is with the School of Electronic Engineering, University of
Electronic Science and Technology of China, Chengdu, Sichuan, China \
610054, (e-mail: huangbwly@126.com).}  \and Hu Zhang\thanks{H. Zhang is with the Department of Applied Mathematics, the Hong Kong Polytechnic University, Hong Kong, China (e-mail: henry6244906@hotmail.com).}}

\begin{document}

\maketitle

\begin{abstract}
The purpose of this paper is to present a theoretic and numerical study of utilizing squeezing and phase shift in coherent feedback control of linear quantum optical systems. A quadrature representation with built-in phase shifters is proposed for such systems. Fundamental structural characterizations of linear quantum optical systems are derived in terms of the new quadrature representation.  These results reveal considerable insights of issue of physical realizability of such quantum systems. The problem of coherent quantum LQG feedback control studied in \cite{NJP09,ZJ11} is re-investigated in depth. Firstly, the optimization methods in \cite{NJP09,ZJ11} are extended to a multi-step optimization algorithm which utilizes ideal squeezers. Secondly, a two-stage optimization approach is proposed on the basis of controller parametrization. Numerical studies show that closed-loop systems designed via the second approach may offer LQG control performance even better than that when the closed-loop systems are in the vacuum state. When ideal squeezers in a close-loop system are replaced by (more realistic) degenerate parametric amplifiers, a sufficient condition is derived for the asymptotic stability of the resultant new closed-loop system; the issue of performance convergence is also discussed in the LQG control setting.
\end{abstract}




\section{Introduction} \label{sec:intro}

Being an important branch of quantum physics, quantum optics has been undergoing an accelerating growth in its applications to emerging quantum
technology as it offers building blocks for constructing quantum computing, communication and metrology devices to realize the dream of quantum
nano-scale technology \cite{Nature09,WM08}. As sound approximations to the fundamental field models in quantum optics, linear quantum optical systems
have been developed in terms of quantum stochastic differential equations (QSDEs) \cite{HP84} (here the word `linear' refers to the linearity of
the Heisenberg equations of motion of system operators), based on which a vast body of measurement-based feedback control methods have been proposed
to achieve such objectives as entanglement preservation, state preparation, error correction, etc. Interested reader may refer to \cite{Bel83,DJ99,WMW02,DF04,vHSM05,BvHJ07,BvHJ09,WM10} for excellent discussions on measurement-based feedback control and their applications to a broad range of quantum optical systems.

An alternative feedback mechanism, \textit{coherent feedback control}, has been proposed recently where measurement is not necessarily involved in feedback loops; instead, quantum information may flow directionally as a (possibly non-commutative) signal (such as a quantum optical electromagnetic field or an injected laser), or directly via a bidirectional physical coupling. The benefits of coherent feedback include (i) preservation of quantum correlation of the whole network, and (ii) high speed \cite[Fig. 1]{IK08}. In \cite{WM94b} an all-optical feedback mechanism is studied for a quantum optical system comprised of two cavities (a source cavity and a driven cavity); By designing appropriate interaction Hamiltonian coupling, the authors were able to obtain squeezed state inside the source cavity. In \cite{YK03} Yanagisawa and Kimura derived closed-loop linear quantum system models consisting of cavities and beamsplitters. A scheme is proposed in \cite{SM06} to produce continuous-wave fields or pulses of polarization-squeezed light via coherent feedback. This proposal is further investigated in \cite{SSM08}. A more general quantum modeling framework is studied in \cite{GJ09}, which is Markovian when channel-to-channel time delay is ignored. This modeling framework is further studied in \cite{GGY08} where transfer functions are obtained for quantum feedback optical networks mediated by beamsplitters. A Hamiltonian formulation of such modeling framework is proposed in \cite{GJ08}. Lately the linear quantum systems framework studied in \cite{YK03} and \cite{GJ08} has been extended to include squeezing components \cite{GJN10}. Input-output maps and transfer functions are defined in this more general framework. In \cite{JNP08} quantum $H^{\infty }$
control of linear quantum stochastic systems is developed, where the resulting controllers can be classical, fully quantum or quantum/classical mixed. Based on the theoretical work developed in \cite{JNP08}, a quantum optical experiment is implemented recently \cite{Mabuchi08}. Because a quantum controller itself is a quantum system, its time evolution must obey Schrodinger's equations. To deal with this fundamental issue, the concept of \emph{physical realizability} is proposed in \cite{JNP08}. Physical realizability is also investigated in \cite{NJD09}, \cite{MP09}, \cite{SP09}, \cite{ZJ11}.

As one of the major methods of modern control, linear quadratic Gaussian (LQG) feedback control has been extended into the quantum domain. In the framework of measurement-based feedback control, quantum LQG feedback control has been investigated in \cite{DJ99,DHJMT00,WD05,EB05,DDJW06,Yam06}, \cite[Sec. 6]{WM10}, etc.. Recently, this problem has been studied in the setting of coherent quantum
feedback networks \cite{NJP09} and \cite{ZJ11}. It turns out that coherent LQG control is more challenging than coherent $H^{\infty }$ control in that control performance and physical realizability are not separable in the LQG coherent feedback control setting. The resulting optimization problem is in general non-convex, and analytical solutions are therefore very difficult to find, if not impossible. A numerical procedure is proposed in \cite{NJP09} which shows that there exists a fully quantum linear controller which offers better closed-loop LQG control performance than classical ones do. A similar procedure is proposed in \cite{ZJ11} which utilizes direct couplings between plants and controllers. Generally speaking, direct coupling is a physical mechanism by which a quantum plant and its quantum controller can exchange energy directly, without necessarily through field connections (cf. \cite[Sec.~II-C and Fig.~4]{JG10}, \cite[Sec.~II-B]{ZJ11}). Direct coupling can be
implemented via nonlinear optical devices like crystals \cite[Fig. 1]{WM94b}.

An ideal squeezer can be modeled as a static Bogoliubov transform \cite[Sec.~II-C]{GJN10}, which is an idealization of a (more realistic) degenerate parametric amplifier (DPA). Ideal squeezers have been used for theoretic study in quantum optics \cite{GZ04,GW09,NJD09,GJN10}. Unfortunately there is as yet no rigorous theoretical justification for their use. Since squeezing devices are prevalent in quantum optics, it is important to address this issue.

Phase modulators are optical elements that manipulate optical waves at both classical and quantum levels; for the latter, they manifest as effects on creation operators of optical modes. Phase shift has been used in a variety of quantum applications, e.g., \cite{KLM01,Mabuchi08,CfP10,petersen11}.

The purpose of this paper is to study quantum optical networks including squeezing components and phase modulators from a control theoretical perspective, aiming at providing systematic control techniques for a wide spectrum of applications. The follows three paragraphs outline the major contributions of the paper.

Firstly, we present a quadrature representation of linear quantum optical systems with built-in phase shifters (Sec.~\ref{sec:models-quadrature}). This representation contains the usual amplitude-phase quadrature representation as a special case. Fundamental algebraic characterizations of such quantum systems are presented in terms of this new representation (Theorems \ref{Thm:phys_real} and \ref{Thm:phys_real2}). These results reveal more insights of the concept physical realizability of open quantum systems first explored in \cite{JNP08}.  Theorems \ref{Thm:phys_real} and \ref{Thm:phys_real2}) are the theoretical basis of the subsequent numerical investigation.

Secondly, quantum LQG coherent feedback control is re-studied in the general framework of linear quantum optical systems including squeezing components and phase shifters presented in Sec.~\ref{systems}. Firstly, we generalize the numerical procedures in \cite{NJP09,ZJ11} by including ideal squeezers and direct couplings in closed-loop systems, and show that performance can be improved considerably (Sec.~\ref{sec:extension}). Secondly, we propose a controller parametrization approach and a two-stage optimization technique to find a coherent feedback controller, ideal squeezers, direct coupling, and/or phase shifters simultaneously. It is shown that appropriate co-design of ideal squeezers, direct coupling, and/or phase shifters can build a closed-loop system which offers considerably good control performance (Secs. \ref{sec:limit} and \ref{sec:limit2}).

Finally, a sufficient condition is derived for the stability of closed-loop systems when ideal squeezers are replaced by DPAs (Theorem \ref{Thm:convergence}). Moreover, a case study is conducted in Sec.~\ref{sec:perf_convg} to demonstrate performance convergence in the LQG feedback control setting. This study hopefully will shorten the gap between the existing theoretical results and their applications, thereby providing experimentalists with some guidance for the choice of parameters of DPAs.

The rest of the paper is organized as follows. In section \ref{systems}, the class of linear quantum optical systems of interest is presented. Section \ref{sec:synthesis} presents the set-up of
closed-loop systems. Section \ref{sec:LQG-synthesis} studies coherent quantum LQG control. Section \ref{sec:squeezer_convergence} studies performance of closed-loop quantum systems when ideal squeezers are replaced by DPAs.  The paper is concluded by Section \ref{sec:conclusion}.

Finally, some words for notation.

\textbf{Notation}. Let $i=\sqrt{-1}$ be the imaginary unit. Given a column
vector of operators $x=[
\begin{array}{ccc}
x_{1} & \cdots & x_{m}%
\end{array}%
] ^{T}$ where $m$ is a positive integer, define $x^{\#}=[
\begin{array}{ccc}
x_{1}^{\ast } & \cdots & x_{m}^{\ast }%
\end{array}%
]^{T}$ where the asterisk $\ast$ indicates Hilbert space adjoint or
complex conjugation. Furthermore, define the doubled-up column vector to be $%
\breve{x}=[
\begin{array}{cc}
x^{T} & \left( x^{\#}\right) ^{T}%
\end{array}%
] ^{T}$. The matrix case can be defined analogously. The symbol $\mathrm{diag%
}_{n}\left( M\right) $ is a block diagonal matrix where the square matrix $M$
appears $n$ times as a diagonal block. Given two matrices $U$, $V\in \mathbb{%
C}^{r\times k}$, a doubled-up matrix $\Delta \left( U,V\right) $ is defined
as $\Delta \left( U,V\right) :=[%
\begin{array}{cccc}
U & V; & V^{\#} & U^{\#}%
\end{array}%
] $. Let $I_{n}$ be an identity matrix. Define $J_{n}=[
\begin{array}{cccc}
0 & I_{n}; & -I_{n} & 0%
\end{array}%
] $ and $\Psi _{n}=\mathrm{diag}(I_{n},-I_{n})$. (The subscript ``$n$'' is
always omitted.) Then for a matrix $X\in \mathbb{C} ^{2n\times 2m}$, define $%
X^{\flat }:=\Psi _{m}X^{\dagger }\Psi _{n}$. Finally, the norm symbol $\Vert
\cdot \Vert $ represents the largest singular value for a constant matrix.

\section{Linear quantum optical systems with squeezing components} \label{systems}

This section introduces systems of interest in the paper. A quadrature representation with embedded phase shifters is introduced, some fundamental relations are presented, an example is used to demonstrate that phase modulation is useful in measurement-based feedback control of linear quantum optical systems.

\subsection{Boson fields}\label{fields}

An $m$-channel free-space light field is described by a vector of annihilation operators $b_{in}(t)=[b_{in,1}(t), \ldots, b_{in,m}(t)]^{T}$ defined on a Fock space \cite{GZ04}, \cite{ZJ11}. These operators satisfy the singular commutation relations
\begin{equation*}
\lbrack b_{in,j}(t),b_{in,k}^{\ast }(t^{\prime })]=\delta _{jk}\delta
(t-t^{\prime }),~~ [b_{in,j}(t),b_{in,k}(t^{\prime })]=[b_{in,j}^{\ast
}(t),b_{in,k}^{\ast }(t^{\prime })]=0.
\end{equation*}%
The operator $b_{in,j}(t)$ ($j=1,\ldots,m$) may be regarded as a quantum stochastic process; in the case where the field is in the vacuum state, this process is quantum white noise. The integrated process $B_{in,j}(t)=\int_{0}^{t}b_{in,j}(\tau)d\tau $ is a quantum Wiener process \cite[Sec.~5.3.5]{GZ04}, which may be used to define quantum stochastic integrals, with associated non-zero Ito
product $dB_{in,j}(t)dB_{in,k}^{\ast }(t)=\delta _{jk}dt$. In this paper we assume that there is no scattering among channels.

\subsection{Open quantum optical systems} \label{sec:models-indirect}

An open quantum optical system $G$ is a collection of $n$ quantum harmonic
oscillators $a(t)=[a_{1}(t),\ldots ,a_{n}(t)]^{T}$ (defined on a Hilbert space) interacting with $m$ boson fields $b_{in}(t)$. Such system can be parameterized by a triple
$\left( I_{m},L,H\right) $. In this triple, the vector operator $L$ is
defined as $L=C_{-}a+C_{+}a^{\#}$, where $C_{-}$ and $C_{+}\in \mathbb{C}^{m\times n}$ and $a=[a_{1},%
\ldots,a_{n}]^{T}$ with $a_{j}$ being the annihilation operator of the $j$th
quantum harmonic oscillator satisfying the canonical commutation relations $%
[a_{j},a_{k}^{\ast }]=\delta _{jk}$. Define a matrix $C=\Delta \left( C_{-},C_{+}\right)$. The observable $H=\frac{1}{2}\breve{a}^\dag\Delta\left(\Omega_{-},\Omega_{+}\right) \breve{a}$
is the initial internal energy of the oscillators, where $\Omega _{-}$ and $\Omega _{+}\in C^{n\times n}$ satisfy $%
\Omega_{-}=\Omega _{-}^{\dag }$ and $\Omega_{+}=\Omega _{+}^{T}$. With these parameters, $G$ on
the composite system-field space can be written as, in Ito form,
\begin{eqnarray}
d\breve{a}(t) &=&-(i\Psi H + \frac{1}{2}C^{\flat }C) \breve{a}(t)dt-C^{\flat}d\breve{B}_{in}\left( t\right) ,~~\breve{a}(0)=\breve{a},  \nonumber
\\
d\breve{B}_{out}\left( t\right) &=&C\breve{a}(t)dt+d\breve{B}_{in}\left(t\right) .  \label{system}
\end{eqnarray}

\subsection{Ideal squeezers}\label{sec:Bog}

In quantum optical experiments, boson fields $b_{in}(t)$ can be squeezed
before they interact with quantum optical systems. Squeezing components can
be approximated by Bogoliubov transformations. Let $S_{-}$ and $S_{+}$ $\in
\mathbb{C}^{m\times m}$. Define $S=\Delta \left( S_{-},S_{+}\right) $. If $S$
satisfies $S^{\flat }S=SS^{\flat }=I_{2m}$, then $S$ is called a \emph{static Bogoliubov transformation}. This paper focuses on a particular class of static Bogoliubov transformations. Let $\Sigma $ be a
real diagonal matrix. Then it can be verified that $S=\exp \left( \Delta \left( 0,\Sigma \right) \right) $ is a static Bogoliubov transformation. In this paper, such type of Bogoliubov transformations are called \textit{%
ideal squeezers}, which have been used in the study of quantum optics \cite{DF04,GZ04}. Clearly, $S$ is an identity matrix if $\Sigma $ is a zero
matrix, that is, there is no squeezing.

\begin{example} \label{Ex:squeezer}
{\rm An ideal squeezer can be regarded as an
idealization of a phase-shifted degenerate parametric amplifier (DPA). A DPA
is an open oscillator that is able to produce squeezed output field. A model
of a DPA is as follows (\cite[page 220]{GZ04}):
\begin{eqnarray*}
d{\breve{a}}(t)&=&-\frac{1}{2}\left[
\begin{array}{cc}
\kappa & -\epsilon \\
-\epsilon & \kappa%
\end{array}%
\right] \breve{a}(t)dt-\sqrt{\kappa }d\breve{B}_{in}(t), \\
d\breve{B}_{out}(t) &=& \sqrt{\kappa}\breve{a}(t)dt + d\breve{B}_{in}(t),
\end{eqnarray*}%
where $\kappa$ and $\epsilon$ are assumed to be real numbers for simplicity and satisfy the inequality $|\epsilon|<\kappa$. In terms of $(I,L,H)$ language, $\Omega _{-}=0$, $\Omega _{+}=\frac{i\epsilon }{2}$, $C_{-}=\sqrt{\kappa }$, and $C_{+}=0$. Let $r=\ln \frac{\kappa +\epsilon }{\kappa -\epsilon }$. Then $S=\exp \left( \Delta \left( 0,r\right) \right) =\Delta \left( \cosh r,\sinh r\right)$ is an ideal squeezer.

Replace $\kappa $ and $\epsilon $ by $\frac{\kappa }{h}$ and $\frac{\epsilon%
}{h}$ respectively where $h>0$, a sequence of DPAs is obtained parameterized by $h$. It can be shown by simple algebra that the sequence of DPA (parameterized by $h$) converges to $e^{i\pi }S$ \emph{pointwisely} as $h \rightarrow 0$.
}
\end{example}


\subsection{Open quantum optical systems with ideal squeezers} \label{sec:system}

If the input fields $b_{in}(t)$ pass through ideal squeezers $S$
before interacting with a collection of open quantum harmonic oscillators,
then the composite system reads
\begin{eqnarray}
d\breve{a}(t) &=&A\breve{a}(t)dt+BSd\breve{B}_{in}\left( t\right) ,~~\breve{a%
}(0)=\breve{a},  \label{model} \\
d\breve{B}_{out}\left( t\right) &=&C\breve{a}(t)dt+Sd\breve{B}_{in}\left(
t\right) ,  \nonumber
\end{eqnarray}%
in which the coefficient matrices are given by
\begin{equation}
A=-\frac{1}{2}C^{\flat }C-i\Psi _{n}\Delta\left(\Omega_{-},\Omega_{+}\right),~~B=-C^{\flat }.  \label{ABCD}
\end{equation}%

\subsection{Quadrature representation of quantum optical systems with embedded phase shifters} \label{sec:models-quadrature}

So far, we have used the annihilation and creation operators $a_{j}$ and $%
a_{j}^{\ast }$ to represent oscillators systems, via the doubled-up notation
$\breve{a}=[a^{T} ~ a^{\dagger }]^{T}$. There is an alternative form,
amplitude-phase quadrature representation, in which all the operators are
observables (self-adjoint operators) and all the corresponding matrices are
real, not imaginary.

\begin{figure}[tbh]
\epsfxsize=2.5in
\par
\epsfclipon
\par
\centerline{\epsffile{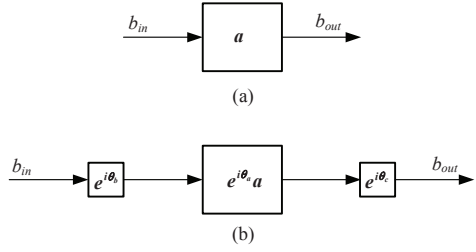}}
\caption{(a) quadrature representation; (b) quadrature representation with embedded phase shifters}
\label{qua_rep}
\end{figure}

Phase shift is a technique which can be used to change phase of a light
beam. Phase modulators are widely used in experimental quantum optics \cite%
{BR04}. In this paper we show that systematic design of phase shifters can
help to improve closed-loop control performance considerably. In this
subsection, we introduce a new type of quadrature representation with
built-in phase shifters.

Let $P_{n}$ be a $2n\times 2n$ permutation matrix which transform a column
vector $d=\left[ d_{1},\ldots ,d_{2n}\right]^T $ to $P_{n}d=\left[%
d_{1},d_{3},d_{5},\ldots ,d_{2},d_{4},\ldots d_{2n}\right]^T $. Let $M_{k}$
be a unitary matrix defined via
\begin{equation*}
M := \frac{1}{\sqrt{2}}\left[\begin{array}{cc}
                               1 & 1 \\
                               -i & i
                             \end{array}
  \right], ~ M_{k} := M\left[
\begin{array}{cc}
e^{i\theta _{k}} & 0 \\
0 & e^{-i\theta _{k}}%
\end{array}%
\right] = \frac{1}{\sqrt{2}}\left[
\begin{array}{cc}
e^{i\theta _{k}} & e^{-i\theta _{k}} \\
-ie^{i\theta _{k}} & ie^{-i\theta _{k}}%
\end{array}%
\right],
\end{equation*}%
where $\theta _{k}$ is real. Introduce%
\begin{equation*}
M_{a}=\mathrm{diag}\left( M_{1},\ldots ,M_{n}\right), M_{b}=\mathrm{diag}%
\left( M_{n+1},\ldots ,M_{n+m}\right), M_{c}=\mathrm{diag}\left(
M_{n+m+1},\ldots ,M_{n+2m}\right) .
\end{equation*}%
Denote matrices
\begin{equation}\label{eq:transformation_matrix}
\Lambda _{a} = P_{n}M_{a}P_{n}^{T}, ~ \Lambda _{b} = P_{m}M_{b}P_{m}^{T}, ~ \Lambda _{c} = P_{m}M_{c}P_{m}^{T}.
\end{equation}
It can be readily verified that $\Lambda _{j}^{\dagger}\Lambda _{j}=\Lambda _{j}\Lambda _{j}^{\dagger }=I, j=a,b,c$. Define a coordinate transform%
\begin{equation}  \label{quadrature}
x:=\Lambda _{a}\breve{a}, ~~ \tilde{B}_{in} :=\Lambda _{b}\breve{B}_{in}, ~~
\tilde{y}:=\Lambda _{c}\breve{B}_{out}.
\end{equation}%
Then in quadrature form, the system $G$ in Eq. (\ref{model}) is converted to
\begin{eqnarray}
dx(t) &=&\tilde{A}x(t)dt+\tilde{B}\tilde{S}d \tilde{B}_{in}(t),~~x(0)=x,
\label{eq:quad-5-dyn} \\
d\tilde{y}(t) &=&\tilde{C}x(t)dt+\Lambda _{c}\Lambda _{b}^{\dagger }\tilde{S}%
d\tilde{B}_{in}(t) ,  \label{eq:quad-5-p}
\end{eqnarray}
where
\begin{equation}
\tilde{A}=\Lambda _{a}A\Lambda _{a}^{\dagger },~{\tilde{B}}=\Lambda
_{a}B\Lambda _{b}^{\dagger },~\tilde{C}=\Lambda _{c}C\Lambda _{a}^{\dagger
},~\tilde{S}=\Lambda _{b}S\Lambda _{b}^{\dagger }.  \label{eq:quad-6}
\end{equation}%

\begin{remark} \label{rem:real} {\rm When all $\theta _{k}$ are $0$, the above quadrature transforms reduce to the unitary transforms used in \cite[Eq. (24)]{ZJ11}} \footnote[1]{In quantum optics, given annihilation and creation operators $a$ and $a^{\dagger}$, amplitude operator is defined to be $a+a^{\dagger}$ while
phase quadrature $-i(a-a^{\dagger})$, cf. \cite[Sec.~4.3.1]{BR04}, \cite[%
Eqs. (2.56), (5.5), (5.6)]{WM08}. Therefore, the quadratures defined by Eq.
(\ref{quadrature}) are the scaled versions of
commonly used quadratures (with the scaling factor $1/\sqrt{2}$). However,
it can be easily seen that these two types of transformations give rise to
the same real matrices.}.
\end{remark}

The above representation contains the amplitude-phase quadrature representation as a special case. In fact, it is the amplitude-phase quadrature representation of the system from $b_{in}$ to $%
b_{out}$ in Fig.~\ref{qua_rep}(b), \emph{not} that in Fig.~\ref{qua_rep}(a). That is, input, output and intra-cavity fields are implicitly assumed to be possibly phase-shifted in this new quadrature representation. This new quadrature representation enables us to choose suitable quadratures which may yield desired (closed-loop) control performance.

\begin{example}
\label{ex:2} {\rm The following linear quantum plant is studied in \cite[%
Sec.~8]{NJP09}:
\begin{eqnarray}
dx(t) &=&\left[
\begin{array}{cc}
0 & \Delta \\
-\Delta & 0%
\end{array}%
\right] x(t)dt+\left[
\begin{array}{cc}
0 & 0 \\
0 & -2\sqrt{\kappa_{1}}%
\end{array}%
\right] d\tilde{u}(t)  \nonumber \\
&&+\left[
\begin{array}{cc}
0 & 0 \\
0 & -2\sqrt{\kappa_{2}}%
\end{array}%
\right] d\tilde{B}_{in,1}(t)+\left[
\begin{array}{cc}
0 & 0 \\
0 & -2\sqrt{\kappa_{3}}%
\end{array}%
\right] d\tilde{B}_{in,2}(t),  \nonumber \\
d\tilde{y}(t) &=&\left[
\begin{array}{cc}
2\sqrt{\kappa_{2}} & 0 \\
0 & 0%
\end{array}%
\right] x(t)dt+d\tilde{B}_{in,1}(t),  \nonumber \\
\tilde{z}(t) &=&x(t)+\tilde{\beta}_{u}(t),  \label{example}
\end{eqnarray}%
where $\Delta =0.1$, $\kappa_{1}=\kappa_{2}=\kappa_{3}=0.01$, $\tilde{\beta}%
_{u}$ is the signal part of $\tilde{u}$ (cf. \cite[Eq. (3)]{JNP08}). The
quantum LQG control problem studied in \cite{NJP09} is to design a
controller that minimizes the following performance index:
\begin{equation}
J_{\infty }=\limsup_{t\rightarrow \infty }\frac{1}{t}\int_{0}^{t}\left%
\langle \tilde{z}^{T}(\tau)\tilde{z}(\tau)\right\rangle d\tau.  \label{LQG_index}
\end{equation}%
That is, the aim of control is to steer the quadrature operators as
close as possible to the origin in phase-space $x$ with a minimum controlling
force $\tilde{\beta}_{u}$ for energy consideration.

Define a quadrature representation%
\begin{equation*}
x=\Lambda _{a}\breve{a},~\mathrm{\tilde{u}}=\Lambda _{b}\breve{B}_{in},~%
\tilde{B}_{in,1}=\Lambda \breve{B}_{in,1},~\tilde{B}_{in,2}=\Lambda \breve{B}%
_{in,2},~\tilde{y}=\Lambda _{c}\breve{B}_{out,1},
\end{equation*}%
where%
\begin{eqnarray*}
\Lambda _{a} &=&\frac{1}{\sqrt{2}}\left[
\begin{array}{cc}
e^{i\theta _{1}} & e^{-i\theta _{1}} \\
-ie^{i\theta _{1}} & ie^{-i\theta _{1}}%
\end{array}%
\right] ,~\Lambda _{b}=\frac{1}{\sqrt{2}}\left[
\begin{array}{cc}
e^{i\theta _{2}} & e^{-i\theta _{2}} \\
-ie^{i\theta _{2}} & ie^{-i\theta _{2}}%
\end{array}%
\right] , \\
\Lambda &=&\frac{1}{\sqrt{2}}\left[
\begin{array}{cc}
1 & 1 \\
-i & i%
\end{array}%
\right] ,~\Lambda _{c}=\frac{1}{\sqrt{2}}\left[
\begin{array}{cc}
e^{i\theta _{3}} & e^{-i\theta _{3}} \\
-ie^{i\theta _{3}} & ie^{-i\theta _{3}}%
\end{array}%
\right] .
\end{eqnarray*}%
Then, in quadrature representation the system is%
\begin{eqnarray}
dx &=&\tilde{A}xdt+\tilde{B}d\tilde{u}+\tilde{B}_{1}d\tilde{B}_{in,1}+\tilde{%
B}_{2}d\tilde{B}_{in,2},  \label{ex:plant} \\
\tilde{y} &=&\tilde{C}_{2}xdt+\tilde{D}_{21}d\tilde{B}_{in,1},  \nonumber \\
\tilde{z} &=&x+\tilde{\beta}_{u},  \nonumber
\end{eqnarray}%
where $\tilde{\beta}_{u}$ is the signal part of $\tilde{u}$, and
\begin{equation*}
\tilde{A} = \left[
\begin{array}{cc}
0 & \Delta \\
-\Delta & 0%
\end{array}%
\right] , ~ \tilde{B}=2\sqrt{\kappa _{1}}\left[
\begin{array}{cc}
-\sin (\theta _{1})\sin (\theta _{2}) & \sin (\theta _{1})\cos (\theta _{2})
\\
\cos (\theta _{1})\sin (\theta _{2}) & -\cos (\theta _{1})\cos (\theta _{2})%
\end{array}%
\right] ,
\end{equation*}
\begin{equation*}
\tilde{B}_{l}=2\sqrt{\kappa _{l}}\left[
\begin{array}{cc}
0 & \sin (\theta _{1}) \\
0 & -\cos (\theta _{1})%
\end{array}%
\right] , ~ \tilde{C}_{2} = 2\sqrt{\kappa _{2}}\left[
\begin{array}{cc}
\cos (\theta _{3})\cos (\theta _{1}) & \cos (\theta _{3})\sin (\theta _{1})
\\
\sin (\theta _{3})\cos (\theta _{1}) & \sin (\theta _{3})\sin (\theta _{1})%
\end{array}%
\right],
\end{equation*}
\begin{equation*}
\tilde{D}_{21}=\left[
\begin{array}{cc}
\cos (\theta _{3}) & -\sin (\theta _{3}) \\
\sin (\theta _{3}) & \cos (\theta _{3})%
\end{array}%
\right] ,~l=1,2.
\end{equation*}
}

{\rm To compare with results in \cite[Sec.~8.2]{NJP09}, only the first
element of $\tilde{y}$ is measured. When $\theta _{1}=\theta _{2}=\theta _{3}=0$, the system reduces to Eq. (38)
in \cite{NJP09}. It has been shown \cite[Sec.~8.2]{NJP09} that a
measurement-based feedback controller yields a closed-loop LQG index $4.8468$. In what follows we study several other cases.

(i). When $\theta _{1}=\theta _{2}=0$ and $\theta _{3}\neq 0$, we
have continuous measurement by homodyne detection \cite{WD05,BR04}.
Numerical study shows that $\theta _{3}=0$ gives rise to an LQG performance $J_{\infty}=4.8468$. That is, there is no improvement.

(ii). When $\theta _{1}=0$, $\theta _{2}=-0.5294$, and $\theta
_{3}=-0.5498$, $J_{\infty}=4.0551$, which is not only better
than $4.8468 $, but also better than $4.1793$ --- the best coherent LQG
control performance obtained in \cite{NJP09}.

(iii). When $\theta _{1}=0.5200$ and $\theta _{2}=\theta _{3}=0$, $3J_{\infty}=.7544$.

(iv). When $\theta _{1}=0.04$, $\theta _{2}=-0.49$, and $\theta_{3}=-0.1$, $J_{\infty} = 3.7388$.

This example demonstrates that measurement-based feedback plus phase shift can yield LQG control performance better than that via coherent feedback studied in \cite{NJP09} and \cite{ZJ11}.
 }
\end{example}

\subsection{Fundamental Relations} \label{sec:funda_reltn}

In this subsection characterizations of open quantum systems are presented.

The following result reveals a fundamental structural relation that the quantum systems under study have.

\begin{theorem} \label{Thm:phys_real}
The real matrices $\tilde{A}$, $\tilde{B}$, and $\tilde{C}$ in Eq. (\ref{eq:quad-6}) satisfy the following equations:%
\begin{equation}  \label{phil_1}
\tilde{A}J_{n}+J_{n}\tilde{A}^{T}+\tilde{B}J_{m}\tilde{B}^{T }=0,
\end{equation}%
\begin{equation}  \label{phil_2}
\tilde{B}=-iJ_{n}\tilde{C}^{T}( \Lambda _{c}\Psi _{m}\Lambda _{b}^{\dagger
}) .
\end{equation}
\end{theorem}

\noindent \textbf{Proof.} It is easy to show that matrices $A$ and $C$ in
Eq. (\ref{ABCD}) satisfy%
\begin{equation*}
\Psi _{n}A+A^{\dag}\Psi _{n}+C^{\dag}\Psi _{m}C=0,
\end{equation*}%
which leads to%
\begin{equation}  \label{phil_3}
\Lambda _{a}\Psi _{n}\Lambda _{a}^{\dagger }\Lambda _{a}A\Lambda
_{a}^{\dagger }+\Lambda _{a}A^{\dagger }\Lambda _{a}^{\dagger }\Lambda
_{a}\Psi _{n}\Lambda _{a}^{\dagger }+\Lambda _{a}C^{\dagger }\Lambda
_{c}^{\dagger }\Lambda _{c}\Psi _{m}\Lambda _{c}^{\dagger }\Lambda
_{c}C\Lambda _{a}^{\dagger }=0.
\end{equation}%
Substituting $\Lambda _{a}\Psi _{n}\Lambda _{a}^{\dagger }=iJ_{n}$ and  $\Lambda _{c}\Psi_{m}\Lambda _{c}^{\dagger }=iJ_{m}$ into Eq. (\ref{phil_3}) yields
\begin{equation} \label{phil_4}
J_{n}\tilde{A}+\tilde{A}^{T }J_{n}+\tilde{C}^{T }J_{m}\tilde{C}=0 .
\end{equation}%
Moreover, by $B=-C^{\flat }$ we have $\Lambda _{a}B\Lambda _{b}^{\dagger }=J_{n}^{\dag }\left( \Lambda _{c}C\Lambda _{a}^{\dagger
}\right) ^{\dagger } ( \Lambda _{c}\Psi _{m}\Lambda _{b}^{\dagger })$. Consequently
\begin{equation}\label{eq:B_tilde}
\tilde{B}=-iJ_{n}\tilde{C}^{T }( \Lambda _{c}\Psi _{m}\Lambda_{b}^{\dagger }) ,
\end{equation}%
which is exactly Eq. (\ref{phil_2}). Moreover, by Eq. (\ref{eq:B_tilde}), $\tilde{C}=-i\Lambda _{c}\Psi _{m}\Lambda _{b}^{\dagger }\tilde{B}^{T }J_{n}$. Substitution of $\tilde{B}$ and $\tilde{C}$ derive above into Eq. (\ref{phil_4}) gives
\begin{eqnarray}
J_{n}\tilde{A}+\tilde{A}^{T }J_{n}+\tilde{C}^{T }J_{m}\tilde{C} &=&J_{n}\tilde{A}+\tilde{A}^{T }J_{n}-J_{n}\tilde{B}\Lambda _{b}\Psi _{m}\Lambda
_{c}^{\dagger }J_{m}\Lambda _{c}\Psi _{m}\Lambda _{b}^{\dagger }\tilde{B}^{T}J_{n} \nonumber
\\
&=&J_{n}\tilde{A}+\tilde{A}^{T }J_{n}-J_{n}\tilde{B}J_{m}\tilde{B}^{T }J_{n} \nonumber
\\
&=&0. \label{eq:temp1}
\end{eqnarray}%
Hence Eq. (\ref{phil_1}) holds. The proof is completed. $\Box$

The following result characterizes solutions to the equations (\ref{phil_1})-(\ref{phil_2}). This result connects equations (\ref{phil_1})-(\ref{phil_2}) with open quantum systems of the form (\ref{system}).

\begin{theorem} \label{Thm:phys_real2}
Given matrices $\Lambda_a,\Lambda_b,\Lambda_c$ defined in Eq. (\ref{eq:transformation_matrix}), there is a unique matric $C$ of the form $C=\Delta(C_-,C_+)$ and infinitely many matrices $A$ of the form $A=-\frac{1}{2}C^{\flat }C-i\Psi _{n}\Delta\left(\Omega_{-},\Omega_{+}\right)$ such that matrices $\tilde{A},\tilde{B},\tilde{C}$ defined via Eq. (\ref{eq:quad-6}) satisfy Eqs. (\ref{phil_1})-(\ref{phil_2}).
\end{theorem}

\noindent \textbf{Proof.} We first show that the matrix $A$ has the specified form.  By Eqs. (\ref{phil_1}) and (\ref{eq:temp1}) we have $J_{n}\tilde{A}+\tilde{A}^{T }J_{n}+\tilde{C}^{T }J_{m}\tilde{C}=0$. As a result, the matrix $\tilde{A}$ is in the form of $\tilde{A} = \frac{1}{2}J_n \tilde{C}^\dag J_m \tilde{C} + J_n R$, where $R$ is an arbitrary real symmetric matrix. Consequently, in terms of $A = \Lambda_a^\dag \tilde{A}\Lambda_a$ and $C = \Lambda_c^\dag \tilde{C}\Lambda_a$ we have $A = -\frac{1}{2}C^\flat C -i\Psi_n  \Lambda_a^\dag R\Lambda_a$. However it is straightforward to show that $\Lambda_a^\dag R\Lambda_a$ is in the form of $\Delta\left(\Omega_{-},\Omega_{+}\right)$. Finally, since $R$ is arbitrary, there are infinitely many $A$ of the form $A=-\frac{1}{2}C^{\flat }C-i\Psi _{n}\Delta\left(\Omega_{-},\Omega_{+}\right)$. Next we show that the matrix $C$ has the specified form. Given a real matrix $\tilde{B}$, some simple algebraic manipulation gives that $-iJ_n \tilde{B}\Lambda_b \Psi_m = [X^\#  ~~ X]$ for some complex matrix $X$. Substituting it into Eq. (\ref{phil_2}) yields $\tilde{C} =\Lambda_c [X ~ X^\#]^T$. Consequently,
\[
C = \Lambda_c^\dag \tilde{C} \Lambda_a = \left[\begin{array}{c}
                             X^T \\
                             X^\dag
                           \end{array}
 \right]\Lambda_a = \Delta(C_-, C_+)
\]
for some matrices $C_-$ and $C_+$. The proof is completed. $\Box$

\begin{remark}{\rm
Eqs (\ref{phil_1}) and (\ref{phil_2}) are forms of \emph{physical realizability} investigated in \cite{JNP08} which reveals fundamental
relations that a quantum system must obey. The derivation of physical realizability in \cite{JNP08} is much mathematically involved. Interested reader may check the proofs
of Theorems 2.1 and 3.4 in \cite{JNP08}. In contrast, the derivation presented here is much simpler as we make use of connections between the annihilation-creation
and quadrature representations. Moreover, Theorem \ref{Thm:phys_real} extends the results in \cite{JNP08} to linear quantum optical systems with built-in phase shifters. Therefore, the results presented here are more general, cf. \cite[Remark 3.6]{JNP08}. These results are used in the numerical study carried out in Sec.~\ref{sec:LQG-synthesis}.
}
\end{remark}

\begin{remark}{\rm
The indication of Theorem \ref{Thm:phys_real2} is two-fold. Firstly, for a given triple $(\Lambda_a,\Lambda_b,\Lambda_c)$, the matrix $C$ is uniquely determined by Eqs. (\ref{phil_1}) and (\ref{phil_2}), while $A$ has free parameters. That is, an open quantum system of the form (\ref{system}) solving Eqs. (\ref{phil_1}) and (\ref{phil_2}) has unique input-output coupling structure while its initial internal Hamiltonian is allowed to be arbitrary. Secondly,  the free choice of triple $(\Lambda_a,\Lambda_b,\Lambda_c)$  provides us with freedom of choosing quantum systems that solve Eqs. (\ref{phil_1}) and (\ref{phil_2}). This advantage is utilized in the numerical study in Sec.~\ref{sec:limit2}.
}
\end{remark}


\section{Closed-loop systems} \label{sec:synthesis}

This section presents the set-up of the closed-loop systems.

\begin{figure}[tbh]
\epsfxsize=2.0in
\par
\epsfclipon
\par
\centerline{\epsffile{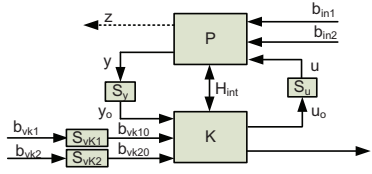}}
\caption{Coherent feedback control}
\label{closed-loop}
\end{figure}

Consider the closed-loop system as shown in Fig.~\ref{closed-loop}, where $P$
is a quantum plant and $K$ a coherent feedback controller to be designed. $S_u,S_y,S_{vK1}, S_{vK2}$ are ideal squeezers to be designed. $b_{in,1}$ and $b_{in,2}$ are quantum noises, $b_{in,1}$ is in vacuum state, while $b_{in,2}$ may have finite energy. $y$ is plant output (output
field channel), $u$ is control input (input field channel), $u_{o}$ is part of the
output of the controller $K$. $z$ is reference output (namely, a performance
variable, which may not be a physical variable, so is designated by a dotted
line), $b_{v_{k1}}$ and $b_{v_{k2}}$ are quantum vacuum noise inputs to the controller $K$. $H_{int}$ in Fig.~\ref{closed-loop} designates the direct
coupling between $P $ and $K$, cf. \cite[Eq. (7)]{ZJ11}.

In terms of quadrature representation introduced in Sec.~\ref%
{sec:models-quadrature}, the quantum plant $P$ is described by
\begin{eqnarray}
dx(t) &=&\tilde{A}x(t)dt+\tilde{B}d\tilde{u}(t)+\tilde{B}_{1}d\tilde{B}%
_{in,1}(t)+\tilde{B}_{2}d\tilde{B}_{in,2}(t),~~x(0)=x,  \nonumber \\
d\tilde{z}(t) &=&\tilde{C}_{1}x(t)dt+\tilde{D}\tilde{u}(t)+\tilde{D}_{12}d%
\tilde{B}_{in,2}(t),  \nonumber \\
d\tilde{y}(t) &=&\tilde{C}_{2}x(t)dt+\tilde{D}_{21}d\tilde{B}_{in,1}(t)+%
\tilde{D}_{22}d\tilde{B}_{in,2}(t),  \label{plant}
\end{eqnarray}%
and the controller by
\begin{eqnarray}
d\xi (t) &=&\tilde{A}_{K}\xi (t)dt+\tilde{B}_{K}d\tilde{y}_{o}(t)+\tilde{B}%
_{K1}d\tilde{B}_{v_{K10}}(t)+\tilde{B}_{K2}d\tilde{B}_{v_{K20}}(t),~~\xi
(0)=\xi ,  \nonumber \\
d\tilde{u}_{o}(t) &=&\tilde{C}_{K}\xi (t)dt+\tilde{D}_{K}d\tilde{B}%
_{v_{K10}}(t).  \label{controller}
\end{eqnarray}%
Note that in the quadrature representation introduced in Sec.~\ref{sec:models-quadrature}, $\tilde{D}_{K}$ in general is not an identity
matrix, as can been seen from Eq. (\ref{eq:quad-5-p}) and the case study in
Sec.~\ref{sec:limit2}.

Define matrices
\begin{equation}
\tilde{A}_{cl}=\left[
\begin{array}{cc}
\tilde{A} & \tilde{B}\tilde{S}_{u}\tilde{C}_{K}+\tilde{B}_{12} \\
\tilde{B}_{K}\tilde{S}_{y}\tilde{C}_{2}+\tilde{B}_{21} & {\tilde{A}}_{K}%
\end{array}%
\right] ,~~\tilde{B}_{cl}=\left[
\begin{array}{c}
\tilde{B}_{2} \\
\tilde{B}_{K}\tilde{S}_{y}\tilde{D}_{22}%
\end{array}%
\right] ,  \label{matrices}
\end{equation}%
\begin{eqnarray*}
\tilde{G}_{cl} &=&\left[
\begin{array}{ccc}
\tilde{B}_{1} & \tilde{B}\tilde{S}_{u}\tilde{D}_{K}\tilde{S}_{vK1} & 0 \\
\tilde{B}_{K}\tilde{S}_{y}\tilde{D}_{21} & \tilde{B}_{K1}\tilde{S}_{vK1} &
\tilde{B}_{K2}\tilde{S}_{vK2}%
\end{array}%
\right] ,~\tilde{C}_{cl}=\left[
\begin{array}{cc}
\tilde{C}_{1} & \tilde{D}\tilde{S}_{u}\tilde{C}_{K}%
\end{array}%
\right] , \\
~~\tilde{H}_{cl} &=&\left[
\begin{array}{ccc}
0 & \tilde{D}\tilde{S}_{u}\tilde{D}_{K}\tilde{S}_{vK1} & 0%
\end{array}%
\right] ,
\end{eqnarray*}%
where $\tilde{B}_{12}$ and $\tilde{B}_{21}$ are for direct coupling,
satisfying $\tilde{B}_{21}=J_{n}\tilde{B}_{12}^{T}J_{n}$ with $\tilde{B}%
_{12} $ being an arbitrary real-valued matrix \cite{ZJ11}. The closed-loop
system with a built-in direct coupling in the quadrature representation is
given by%
\begin{eqnarray}
\left[
\begin{array}{c}
dx(t) \\
d\xi (t)%
\end{array}%
\right] &=&\tilde{A}_{cl}\left[
\begin{array}{c}
x(t) \\
\xi (t)%
\end{array}%
\right] dt+\tilde{B}_{cl}d\tilde{B}_{in,2}(t)+\tilde{G}_{cl}\left[
\begin{array}{c}
d\tilde{B}_{in,1}(t) \\
d\tilde{B}_{v_{K1}}(t) \\
d\tilde{B}_{v_{K2}}(t)%
\end{array}%
\right] ,  \label{cl0} \\
d\tilde{z}(t) &=&\tilde{C}_{cl}\left[
\begin{array}{c}
x(t) \\
\xi (t)%
\end{array}%
\right] dt+\tilde{D}_{12}d\tilde{B}_{in,2}(t)+~\tilde{H}_{cl}\left[
\begin{array}{c}
d\tilde{B}_{in,1}(t) \\
d\tilde{B}_{v_{K1}}(t) \\
d\tilde{B}_{v_{K2}}(t)%
\end{array}%
\right] .  \nonumber
\end{eqnarray}

\section{LQG synthesis} \label{sec:LQG-synthesis}

In the previous section we have added ideal squeezers and phase shifters to closed-loop plant-controller systems. It is natural to investigate if they are helpful for the purpose of control. As mentioned in the Introduction section, coherent LQG quantum feedback control is as yet an outstanding problem, in this section we would like to re-study this problem in order to illustrate the usefulness of the more general framework presented in the previous sections. Example \ref{ex:2} has shown that measurement-based feedback plus phase shifters can outperform the best coherent feedback control performances derived in \cite{NJP09} and \cite{ZJ11}. We show in this section that coherent feedback can achieve much
better performance if ideal squeezers, phase modulators and direct coupling are
appropriately designed in coherent feedback control.

\subsection{Set-up} \label{sec:setup}

In this subsection, we formulate the coherent quantum LQG feedback control
problem.  This problem has been investigated in \cite{NJP09} and \cite{ZJ11}. Similar problems have been investigated in \cite{DJ99}, \cite{DHJMT00}, \cite{WD05}, \cite{EB05}, \cite%
{DDJW06}, \cite{Yam06}, \cite[Sec.~6]{WM10}, etc., in the framework of
measurement-based feedback control. As mentioned in Example \ref{ex:2}, the aim of
control is to minimize the variances of amplitude and phase quadratures with
a minimum controlling force for energy consideration.

The following assumptions are standard (cf. \cite{NJP09}).

\begin{description}
\item[A1.] $\tilde{D}_{12}$ in the plant (\ref{plant}) is zero.

\item[A2.] The quantum noise inputs $b_{in,1} (t)$ and $b_{in,2} (t)$ in the
plant (\ref{plant}) are in vacuum state.

\item[A3.] The plant is initially in Gaussian state.

\item[A4.] The controller is initially in vacuum state.
\end{description}

Moreover, for simplicity, we assume

\begin{description}
\item[A5.] The performance variable is $\tilde{z}(t)=x(t)+\tilde{\beta}%
_{u_o} (t)$, where $\tilde{\beta}_{u_o} (t)$ is the signal part of $u_o (t)$.
\end{description}

Under these assumptions, suppose that $\tilde{A}_{cl}$ in Eq. (\ref{matrices}%
) is Hurwitz, then a positive definite matrix $\mathcal{P}_{LQG}$ is the
(unique) solution to the following Lyapunov equation%
\begin{equation}
\tilde{A}_{cl}\mathcal{P}_{LQG}+\mathcal{P}_{LQG}\tilde{A}_{cl}^{T}+\left[
\begin{array}{cc}
\tilde{B}_{cl} & ~\tilde{G}_{cl}%
\end{array}%
\right] \left[
\begin{array}{cc}
\tilde{B}_{cl} & ~\tilde{G}_{cl}%
\end{array}%
\right] ^{T}=0.  \label{eq:lyap}
\end{equation}%
As is shown in \cite{NJP09}, the closed-loop LQG performance index $%
J_{\infty }$ in (\ref{LQG_index}) can be computed via
\begin{equation}
J_{\infty }=Tr\left( \tilde{C}_{cl}\mathcal{P}_{LQG}\tilde{C}%
_{cl}^{T}\right).  \label{Eq:J_LQG}
\end{equation}
As we wish to minimize the variances of amplitude and phase quadratures, to
conform to footnote 1 in Sec.~\ref{sec:models-quadrature}, we use the following quadrature transformation
\begin{equation}
x = \left[
\begin{array}{cc}
1 & 1 \\
-i & i%
\end{array}%
\right]\left[
\begin{array}{c}
a \\
a^*%
\end{array}%
\right].
\end{equation}

In what follows we first study an extreme case.  Denote
\begin{equation}
\tilde{C}_{K}^{T}\tilde{C}_{K}=\left[
\begin{array}{cc}
c_{1} & c_{2} \\
c_{2} & c_{3}%
\end{array}%
\right] .  \label{eq:CC}
\end{equation}

\begin{theorem} \label{thm:vacuum}
Assume that in steady state both the plant and the controller are in the vacuum state, then $J_{\infty }= 2+(c_{1}+c_{3})$.
\end{theorem}

 \emph{Proof.}  Firstly, it is easy to show that
\begin{equation}
\left\langle \tilde{z}^{T}(s)\tilde{z}(s)\right\rangle =\left\langle
x^{T}(s)x(s)\right\rangle +\left\langle x^{T}(s)\tilde{C}_{K}\xi
(s)\right\rangle +\left\langle \xi^{T}(s)\tilde{C}_{K}^{T}x(s)\right\rangle
+\left\langle \xi^{T}(s)\tilde{C}_{K}^{T}\tilde{C}_{K}\xi (s)\right\rangle .
\label{z4}
\end{equation}%
Secondly, note that
\begin{equation}
x^{T}(s)x(s)=4a^{\ast }(s)a(s)+2,  \label{eq:x}
\end{equation}
and
\begin{equation}
\left\langle \xi (s)^{T}\tilde{C}_{K}^{T}\tilde{C}_{K}\xi (s)\right\rangle
=\left\langle c_{1}q_{\xi }^{2}(s)+c_{2}q_{\xi }(s)p_{\xi }(s)+p_{\xi
}(s)q_{\xi }(s)+c_{3}p_{\xi }^{2}(s)\right\rangle ,  \label{eq:xi}
\end{equation}%
where%
\begin{eqnarray*}
q_{\xi}^{2} &=& a_{K}^{2}+1+2a_{K}^{\ast }a_{K}+(a_{K}^{\ast })^{2}, \ \ p_{\xi}^{2} = -\left( a_{K}^{2}-1-2a_{K}^{\ast }a_{K}+(a_{K}^{\ast })^{2}\right) , \\
q_{\xi }p_{\xi }+p_{\xi }q_{\xi } &=& -2i(a_{K}^{2}-(a_{K}^{\ast })^{2}).
\end{eqnarray*}
Substituting Eqs. (\ref{eq:x}) and (\ref{eq:xi}) into Eq. (\ref{z4}) yields%
\begin{eqnarray}
\left\langle \tilde{z}^{T}(s)\tilde{z}(s)\right\rangle &=&-2ic_{2}\left\langle a_{K}^{2}(s)-(a_{K}^{\ast }(s))^{2}\right\rangle
+(c_{1}-c_{3})\left\langle a_{K}^{2}(s)+(a_{K}^{\ast }(s))^{2}\right\rangle \label{z5}
 \\
&&+\left\langle x^{T}(s)\tilde{C}_{K}\xi (s)\right\rangle +\left\langle \xi(s)^{T}\tilde{C}_{K}^{T}x(s)\right\rangle +2+(c_{1}+c_{3}).  \nonumber
\end{eqnarray}
As both the plant and the controller are in the vacuum state in steady state,
\begin{equation}\label{temp_1}
\left\langle a(s)^{\ast }a(s)\right\rangle=0, \left\langle
a_{K}^{2}(s)\right\rangle =\left\langle (a_{K}^{\ast }(s))^{2}\right\rangle
=0, \left\langle x^{T}(s)\tilde{C}_{K}\xi (s)\right\rangle =\left\langle
\xi^{T}(s)\tilde{C}_{K}^{T}x(s)\right\rangle =0.
\end{equation}
Substituting Eq.~(\ref{temp_1}) into Eq. (\ref{z5}) leads to $\left\langle \tilde{z}^{T}(s)\tilde{z}(s)\right\rangle = 2+(c_{1}+c_{3})$. Consequently, $J_{\infty }= 2+(c_{1}+c_{3})$. The proof is completed. $\Box$

\begin{remark} \label{rem:limit}
{\rm
It is worth pointing out that $J_{\infty }$ derived above is a result of the Heisenberg's uncertainty principle. Clearly, $J_{\infty }$ is independent of plant parameters.
}
\end{remark}

\begin{remark}\label{rem:lqg}
{\rm The plant studied in Example \ref{ex:2} can model an optical cavity interacting with three input fields. If the term $\tilde{\beta}_{u}$ is ignored in Eq. (\ref{example}) by setting $C_K = 0$, then it is easy to see that the performance index defined in Eq. (\ref{LQG_index}) means minimization of the number of photons in the cavity. Theorem \ref{thm:vacuum} shows that the minimal value is 2. However, the plant in this example is only marginally stable, so it has to be controlled. Therefore $C_K$ cannot be zero and hence $J_{\infty }= 2+(c_{1}+c_{3}) > 2$.
}
\end{remark}

As direct coupling can add new energy channel between the plant and its controller, squeezers can modify the input-output structure of the plant, and phase modulators can adjust relative phases of signals in the closed loop, in this sequel we seek to study the following problem.

\begin{problem}
{\rm
Is it possible to design a coherent feedback controller, direct coupling,
ideal squeezers and phase modulators simultaneously such that the resulting
closed-loop system performance better than $2+(c_{1}+c_{3})$?
}
\end{problem}

The measurement-based feedback studied in Example \ref{ex:2} fails to
provide an affirmative answer to the above problem. Moreover, the coherent
controllers proposed in \cite{NJP09} and \cite{ZJ11} fail too.


\subsection{An Extension of algorithms in \cite{NJP09,ZJ11}} \label{sec:extension}

In this subsection we re-investigate the example studied in \cite{NJP09,ZJ11}. It
turns out that an appropriate choice of ideal squeezers and direct couplings can
improve the performance of the fully quantum controller
considerably. This case study embodies the essential ingredients of quantum
LQG controller design, thus algorithms for more general cases can be
developed without additional conceptual difficulty.

In the following we present a numerical optimization algorithm.

Let $2$ by $2$ real matrices $Z_{x_{1},1}^{T},\ldots,Z_{x_{12},1}^{T},Z_{v_{1},1}^{T},\ldots ,Z_{v_{14},1}^{T}$ be a set of decision variables matrices. Denote $V=\left[ I;Z_{x_{1},1}^{T};\cdots ;Z_{x_{12},1}^{T};Z_{v_{1},1}^{T};\cdots;Z_{v_{14},1}^{T}\right] ^{T}$. Define a symmetric matrix $Z=VV^{T}$. For ease of presentation, denote
\begin{eqnarray*}
Z_{x} &=&\left[
\begin{array}{ccc}
Z_{x_{2},1} & Z_{x_{3},1} & Z_{x_{4},1}%
\end{array}%
\right] ,~~B_{w}^{^{\prime }}=\left[
\begin{array}{ccc}
\tilde{B}_{2} & \tilde{B}\tilde{S}_{u}\tilde{S}_{vK1} & 0%
\end{array}%
\right] , \\
C_{w}^{^{^{\prime \prime }}} &=&\left[
\begin{array}{ccc}
\tilde{S}_{vK1} & 0 & 0 \\
0 & \tilde{S}_{vK2} & 0 \\
0 & 0 & \tilde{S}_{y}%
\end{array}%
\right] \left[
\begin{array}{c}
0 \\
0 \\
\tilde{C}_{2}%
\end{array}%
\right] ,~~D_{w}^{^{^{\prime \prime }}}=\left[
\begin{array}{ccc}
\tilde{S}_{vK1} & 0 & 0 \\
0 & \tilde{S}_{vK2} & 0 \\
0 & 0 & \tilde{S}_{y}%
\end{array}%
\right] \left[
\begin{array}{ccc}
0 & I & 0 \\
0 & 0 & I \\
\tilde{D}_{21} & 0 & 0%
\end{array}%
\right] .
\end{eqnarray*}%
Construct the following linear matrix inequalities
\begin{equation*}
{\small \left[
\begin{array}{ccc}
\tilde{A}Z_{x_{6},1}\mathbf{+}\tilde{B}\tilde{S}_{u}Z_{x_{5},1}\mathbf{+}(%
\tilde{A}Z_{x_{6},1}\mathbf{+}\tilde{B}\tilde{S}_{u}Z_{x_{5},1})^{T} &
Z_{x_{1},1}^{T}+\tilde{A} & B_{w}^{^{\prime }} \\
Z_{x_{1},1}\mathbf{+}\tilde{A}^{T} & Z_{x_{7},1}\tilde{A}+Z_{x}C_{w}^{^{^{%
\prime \prime }}}+(Z_{x_{7},1}\tilde{A}+Z_{x}C_{w}^{^{^{\prime \prime
}}})^{T} & Z_{x_{7},1}B_{w}^{^{\prime }}+Z_{x}D_{w}^{^{^{\prime \prime }}}
\\
(B_{w}^{^{\prime }})^{T} & (Z_{x_{7},1}B_{w}^{^{\prime
}}+Z_{x}D_{w}^{^{^{\prime \prime }}})^{T} & -I%
\end{array}%
\right] }
\end{equation*}%
\begin{equation}
\hspace{-4.1cm}{\small +\left[
\begin{array}{ccc}
\tilde{B}_{12}+\tilde{B}_{12}^{T} & \left( Z_{x_{12},1}\mathbf{+}Z_{x_{7},1}%
\tilde{B}_{12}\right) ^{T} & 0 \\
Z_{x_{12},1}\mathbf{+}Z_{x_{7},1}\tilde{B}_{12} &
Z_{x_{11},1}+Z_{x_{11},1}^{T} & 0 \\
0 & 0 & 0%
\end{array}%
\right] <0,}  \label{lmi:v1}
\end{equation}%
\begin{equation}
\left[
\begin{array}{ccc}
Z_{x_{6},1} & I & \left( \tilde{C}_{1}Z_{x_{6},1}+\tilde{D}%
_{12}Z_{x_{5},1}\right) ^{T} \\
I & Z_{x_{7},1} & \tilde{C}_{1}^{T} \\
\tilde{C}_{1}Z_{x_{6},1}+\tilde{D}_{12}Z_{x_{6},1} & \tilde{C}_{1} & Q%
\end{array}%
\right] >0,  \label{lmi:v2}
\end{equation}%
\begin{equation}
Tr(Q)<\gamma ,  \label{trace}
\end{equation}%
a rank constraint%
\begin{equation}
rank(Z)\leq 2,  \label{rank}
\end{equation}%
and a set of additional constraints%
\begin{equation}
\begin{array}{ll}
Z\geq 0, & Z_{x_{8},1}+Z_{x_{9},x_{10}}=0, \\
Z_{0,0}-I=0, & Z_{v_{4},1}-Z_{x_{7},1}\tilde{B}_{2}\tilde{S}_{u}=0, \\
Z_{1,x_{6}}-Z_{x_{6},1}=0, & Z_{v_{6},1}-Z_{x_{8},x_{5}}=0, \\
Z_{1,x_{7}}-Z_{x_{7},1}=0, & Z_{v_{7},1}-Z_{x_{8},x_{6}}=0, \\
Z_{v_{9},1}-Z_{x_{7},x_{6}}=0, & Z_{v_{8},1}-Z_{x_{1},x_{8}}=0, \\
Z_{v_{10},1}-Z_{v_{4},v_{6}}=0, & Z_{x_{10},1}+Z_{1,x_{10}}=0, \\
Z_{v_{11},1}-Z_{v_{5},v_{7}}=0, & Z_{v_{13},1}-Z_{v_{2},x_{3}}=0, \\
Z_{v_{12},1}-Z_{v_{1},x_{2}}=0, & Z_{v_{14},1}-Z_{v_{3},x_{4}}=0, \\
Z_{v_{1},1}-Z_{x_{2},1}J_{N_{v_{_{K1}}}}=0, &
Z_{v_{2},1}-Z_{x_{3},1}J_{N_{v_{_{K2}}}}=0, \\
Z_{x_{9},1}-J_{N_{\xi }}+Z_{v_{9},1}=0, & Z_{v_{3},1}-Z_{x_{4},1}J_{N_{y}}=0,
\\
Z_{v_{5},1}-Z_{x_{4},1}\tilde{C}_{2}-Z_{x_{7},1}\tilde{A}=0, &  \\
Z_{x_{12},1}=Z_{x_{11},x_{6}}, & Z_{x_{11},1}=Z_{x_{9},1}\tilde{B}_{21}.%
\end{array}
\label{28_v2}
\end{equation}%
Two equality constraints for physical realizability:
\begin{equation}
-Z_{v_{8},1}+Z_{v_{8},1}^{T}+Z_{v_{11},1}-Z_{v_{11},1}^{T}+Z_{v_{10},1}-Z_{v_{10},1}^{T}+Z_{v_{12},1}+Z_{v_{13},1}+Z_{v_{14},1}=0,
\label{29_v2_a}
\end{equation}%
and%
\begin{equation}
Z_{x_{2},1}-Z_{v_{6},1}J_{N_{v_{_{K1}}}}=0.  \label{29_v2_b}
\end{equation}%
The quantum LQG control algorithm is given below.

\textit{Initialization}. Set $\tilde{B}_{12}=0$, $\tilde{S}_{u}=I$, $\tilde{S%
}_{y}=I$, $\tilde{S}_{v_{K1}}=I$, and $\tilde{S}_{v_{K2}}=I$.

\textit{Step 1}. For fixed $\tilde{B}_{12}$, $\tilde{S}_{u}$, $\tilde{S}_{y}$%
, $\tilde{S}_{v_{K1}}$, and $\tilde{S}_{v_{K2}}$, given $\gamma >0$, employ
a semidefinite programming (SDP) approach to solve the feasibility problem
with constraints Eqs. (\ref{lmi:v1}) -- (\ref{29_v2_b}) in which
decision variables are $Z$ and $Q$.

\textit{Step 2. } Pertaining to \textit{Step 1}, use some optimization
procedure to find direct coupling parameters $\tilde{B}_{12 }$.

\textit{Step 3. } Pertaining to \textit{Step 2}, use some optimization
procedure to find squeezers $\tilde{S}_{u}$, $\tilde{S}_{v_{K1}}$ and $%
\tilde{S}_{v_{K2}}$. Then go to \textit{Step 1}.

\textit{Step 1} can be solved using an algorithm similar to that proposed in
\cite{NJP09} which is based on LMIRank (\cite{OHM06}), SeDuMi (\cite{SeDuMi}%
), and Yalmip (\cite{Lo04}), while in \textit{Steps 2} and \textit{3} only a
few decision variables are involved, hence they can be easily solved via many
general-purpose optimization algorithms.

If the above optimization problem is solvable, controller parameters of $K$
can be obtained via $\tilde{A}_{K} = Z_{x_{1},1}, ~ \tilde{B}_{K1}= Z_{x_{2},1}, ~ \tilde{B}%
_{K2}=Z_{x_{3},1}, ~ \tilde{B}_{K} =Z_{x_{4},1}, ~ \tilde{C}_{K}=
Z_{x_{5},1}$.

In what follows we present the solution of the preceding numerical
algorithm. After initialization, solving \textit{step 1} yields a controller
$K$ with parameters
\begin{equation*}
\tilde{A}_{K}=\left[
\begin{array}{cc}
0.0251 & -0.3787 \\
0.0665 & -0.2121%
\end{array}%
\right] ,\tilde{B}_{K}=\left[
\begin{array}{cc}
1.0273 & -0.1964 \\
0.8235 & -0.0492%
\end{array}%
\right] ,\tilde{B}_{K1}=\left[
\begin{array}{cc}
0.1125 & -0.5992 \\
0.1504 & -0.1284%
\end{array}%
\right] ,
\end{equation*}%
\begin{equation*}
\tilde{B}_{K2}=10^{-10}\left[
\begin{array}{cc}
0.6008 & -0.3049 \\
0.1938 & -0.2273%
\end{array}%
\right] ,~~\tilde{C}_{K}=\left[
\begin{array}{cc}
0.1284 & -0.5993 \\
0.1506 & -0.1126%
\end{array}%
\right] ,~~\tilde{D}_{K}=I.
\end{equation*}%
which yields $J_{\infty }=4.1787$. Implement \textit{step 2} produces a
direct coupling with
\begin{equation*}
\tilde{B}_{12}=\left[
\begin{array}{cc}
1.2 & -9 \\
0.72 & 0.36%
\end{array}%
\right] \times 10^{-3},~~~\tilde{B}_{21}=\left[
\begin{array}{cc}
-0.36 & -9 \\
0.72 & -1.2%
\end{array}%
\right] \times 10^{-3},
\end{equation*}%
The it is found that $J_{\infty }=3.9995$. Implementing \textit{step 3}
generates squeezers
\begin{equation*}
\tilde{S}_{u}=\tilde{S}_{y}=I,~\tilde{S}_{v_{K1}}=\left[
\begin{array}{cc}
1.5876 & 0 \\
0 & 0.6299%
\end{array}%
\right] ,~\tilde{S}_{v_{K2}}=\left[
\begin{array}{cc}
1.8076 & 0 \\
0 & 0.5532%
\end{array}%
\right] .
\end{equation*}%
It can be verified that in this case $J_{\infty }=3.8312$. This is, by
adding appropriate squeezers LQG performance is further improved. With
direct coupling and squeezers obtained above, run \textit{step 1} again to
yield a controller $K$ with parameters
\begin{equation*}
\begin{array}{ll}
\tilde{A}_{K}=\left[
\begin{array}{cc}
0.0108 & -0.4819 \\
0.0353 & -0.2116%
\end{array}%
\right] ,~~\tilde{B}_{K}=\left[
\begin{array}{cc}
1.3697 & -0.4995 \\
0.7232 & -0.1559%
\end{array}%
\right] , \\
\tilde{B}_{K1}=\left[
\begin{array}{cc}
0.2996 & -0.7662 \\
0.1427 & -0.1876%
\end{array}%
\right] , ~~
\tilde{B}_{K2}=10^{-14}\left[
\begin{array}{cc}
0.9902 & 0.6591 \\
0.4085 & -0.0584%
\end{array}%
\right] , \\
\tilde{C}_{K}=\left[
\begin{array}{cc}
0.1876 & -0.7662 \\
0.1427 & -0.2996%
\end{array}%
\right] ,~~\tilde{D}_{K}=I.
\end{array}
\end{equation*}%
and the resulting LQG performance is $J_{\infty }=3.7464$.

\subsection{Controller parametrization and a numerical optimization approach - without phase shifters}\label{sec:limit}

The algorithm in the preceding subsection is an extension of those in \cite%
{NJP09,ZJ11}, however such algorithms suffer from severe
limitation: they are nonlinear and non-convex optimization procedures
involving many decision variables. Consequently, it is
very challenging to use such algorithms to optimize over all controller
parameters, direct couplings, ideal squeezers, and phase shifts
simultaneously.

From Sec.~\ref{sec:models-indirect} we see that controllers can be
parameterized by $C_-$, $C_+$, $\Omega_-$, $\Omega_+$, from Sec.~\ref%
{sec:Bog} we see that each ideal squeezer can be parametrized by a real
number; and direct coupling and phase shifters involve just a couple of
parameters. Moreover, with such parametrization physical realizability is
naturally satisfied. This enlightens us to seek for controller design
methods via optimization over these parameters. Inspired by the study in
\cite{CZ03}, in this and the next sections we investigate a new optimization
approach on the basis of controller parametrization and a two-stage
optimization algorithm.

The idea of controller parametrization is simple. The coherent feedback
controller to be constructed is in the form of Eq. (\ref{controller}), that
is, a quantum harmonic oscillator interacting with three input fields.
According to Sec.~\ref{sec:models-indirect}, there are 13 parameters to be
optimized. Direct coupling requires 4 parameters, while each of the four squeezers
needs one parameter respectively. Therefore, we will optimize over 21
parameters simultaneously. Based on this parametrization, a two-stage optimization is constructed which
is outlined as follows.

In Stage one, a genetic algorithm optimizes over the cost function (\ref%
{Eq:J_LQG}) subject to the constraint that the closed-loop matrix $\tilde{A}_{cl}$ is Hurwitz.

Note that physical realizability is satisfied naturally in terms of
parametrization. The genetic algorithm performs a global search over the
parameter space to find a minimal solution \cite{Gol89}.

In Stage two, a local search is performed around the minimal solution
derived from Stage one. Around this solution, the closed-loop matrix $\tilde{%
A}_{cl}$ is always Hurwitz, hence the optimization problem boils down to
minimizing the cost function (\ref{Eq:J_LQG}) subject to the constraint (\ref%
{eq:lyap}). Here, the cost function and constraints are all continuously
differentiable. Consequently, this optimization problem can be solved by
means of many standard optimization algorithms, cf. \cite{Matlab03}.

\begin{remark}
In contract to the numerical optimization procedure presented in \cite{NJP09,ZJ11} and the previous subsection, the above optimization problem
has a parameter space of much smaller dimension. Moreover, genetic algorithms
usually provide a reasonably good initial solution \cite{Gol89}.
\end{remark}

The preceding numerical algorithm turns out to be very effective. It
provides the following system parameters
\begin{equation*}
\begin{array}{ll}
C_{-}=\left[
\begin{array}{c}
-0.0136 + 0.0857i \\
-0.0473 - 0.3509i \\
2.7099 +19.4445i%
\end{array}%
\right] , ~ C_{+}=\left[
\begin{array}{c}
0.0136 + 0.0857i \\
0.0286 - 0.2251i \\
0.0763 - 0.3437i%
\end{array}%
\right] ,\\
\Omega _{-}=0.9768, ~ \Omega _{+}=-2.4874 - 0.3771i,
\end{array}
\end{equation*}
Therefore, the matrices of controller $K$ are
\[
\tilde{A}_{K}=\left[
\begin{array}{cc}
-193.0685 & 3.4642 \\
1.5106 & -192.3144%
\end{array}%
\right] , ~ \tilde{B}_{K}=\left[
\begin{array}{cc}
-2.6336 & -19.7882 \\
19.1008 & -2.7862%
\end{array}%
\right] ,
\]
\[
\tilde{B}_{K1}=\left[
\begin{array}{cc}
0.0272 & 0.0000 \\
0.1715 & 0.0000%
\end{array}%
\right], ~
\tilde{B}_{K2}=\left[
\begin{array}{cc}
0.0759 & 0.1258 \\
-0.5760 & 0.0187%
\end{array}%
\right],
\]
\[
\tilde{C}_{K}=\left[
\begin{array}{cc}
0 & 0 \\
0.1715 & -0.0272%
\end{array}%
\right] , ~ \tilde{D}_K = I .
\]
The matrices for direct coupling are
\begin{equation*}
\tilde{B}_{12}=\left[
\begin{array}{cc}
151.1269 & 0.0621 \\
1.3904 & 123.8024%
\end{array}%
\right] ,~~\tilde{B}_{21}=\left[
\begin{array}{cc}
-123.8024 & 0.0621 \\
1.3904 & -151.1269%
\end{array}%
\right] .
\end{equation*}%
Finally ideal squeezers are
\begin{eqnarray*}
\tilde{S}_{u} &=&\left[
\begin{array}{cc}
230.3001 & 0 \\
0 & 0.0043%
\end{array}%
\right] ,~\tilde{S}_{v_{K1}}=\left[
\begin{array}{cc}
0.0972 & 0 \\
0 & 10.2920%
\end{array}%
\right] , \\
~\tilde{S}_{v_{K2}} &=&\left[
\begin{array}{cc}
0.5163 & 0 \\
0 & 1.9367%
\end{array}%
\right] , ~ \tilde{S}_{y}=\left[
\begin{array}{cc}
1.1253 & 0 \\
0 & 0.8887%
\end{array}%
\right].
\end{eqnarray*}%
The resulting LQG performance is $2.0004$, which is much better than that
obtained in Sec.~\ref{sec:extension}.

Finally, by Eq. (\ref{eq:CC}),  $2+(c_{1}+c_{3})=2.03013 > 2.0004$. That is, this closed-loop system offers performance even better than the vacuum case given by Theorem \ref{thm:vacuum}.

\subsection{Controller parametrization and a numerical optimization approach - with phase shifters} \label{sec:limit2}

In this subsection we study whether the LQG performance can be further
improved if phase modulators are added. According to the study in the
previous section, the controller, direct coupling and ideals squeezers
obtained there already yields a performance which is better than that in the vacuum case, thus adding phase modulators will not improve performance
considerably.

As shown in Example \ref{ex:2}, the parametrization of the quantum plant
requires three phase values, say $\theta _{1}$, $\theta _{2}$ and $\theta
_{3}$. As far as the controller is concerned, we denote by $\theta _{4}$, $%
\theta _{5}$, and $\theta _{6}$ the phase shifters of the intra-cavity
mode of the controller, the traveling fields $y_o$ and $u_o$ respectively.
That is, there are 6 more optimization variables than the optimization
problem studied in Sec.~\ref{sec:limit}.

Using a similar two-stage optimization procedure as that in Sec.~\ref%
{sec:limit} we find $\theta_1 = 0.2711$, $\theta_2 = -1.0493\times 10^{-4}$,  $\theta_3 =
-0.30233$, $\Omega_- = 3.8780$, $\Omega_+ = 2.7264 + 0.3709i$ and
\[
C_- =10^{2}\times\left[
\begin{array}{c}
0.0010 + 0.0032i \\
-0.0178 + 4.5222i \\
0.0567 + 0.0117i%
\end{array}%
\right], ~ C_+ = \left[
\begin{array}{c}
0.0474 + 0.3290i \\
1.7123 - 6.7882i \\
0.5665 + 0.2253i%
\end{array}%
\right].
\]
The matrices for direct coupling are
\begin{equation*}
\tilde{B}_{12}=10^{5}\times\left[
\begin{array}{cc}
1.3774 & 0.002003 \\
0.005402 & 0.4648%
\end{array}%
\right] ,~~\tilde{B}_{21}=10^{5}\times\left[
\begin{array}{cc}
-0.4648 & 0.002003 \\
0.005402 & -1.3774%
\end{array}%
\right].
\end{equation*}
Finally, the ideal squeezers are
\begin{eqnarray*}
\tilde{S}_{u} &=&10^{3}\times\left[
\begin{array}{cc}
9.2358 & 0 \\
0 & 0.0000001083%
\end{array}%
\right] ,~\tilde{S}_{v_{K1}}=\left[
\begin{array}{cc}
0.4808 & 0 \\
0 & 2.0798%
\end{array}%
\right] , \\
~\tilde{S}_{v_{K2}} &=&\left[
\begin{array}{cc}
1.7476 & 0 \\
0 & 0.5722%
\end{array}%
\right] , ~ \tilde{S}_{y}=\left[
\begin{array}{cc}
1.5537 & 0 \\
0 & 0.6436%
\end{array}%
\right] .
\end{eqnarray*}
The resulting LQG performance is $2.0000008422$, which is slightly better
than $2.0004$.

\begin{remark}
The big difference between parameters of controller, direct coupling and
ideal squeezers obtained in the above two subsections reveals the non-convex
nature of the underlying optimization problem. This in turn reveals the
necessity of simultaneous optimization.
\end{remark}

\section{Performance convergence}\label{sec:squeezer_convergence}
The preceding section has shown that ideal squeezers are very useful in controller design for linear quantum optical systems. In the practice of quantum optics, ideal squeezers are approximations of
more realistic degenerate parametric amplifiers (DPAs). Therefore, it is desirable and practically important to check how closed-loop quantum LQG performance changes when ideals squeezers are replaced by DPAs. In this section, we focus on closed-loop stability and closed-loop LQG performance convergence.

\subsection{Asymptotic stability}\label{sec:appro}

The concept of stability is discussed in \cite[Sec.~III]{ZJ11}. In
particular, for the linear quantum systems discussed in this paper,
asymptotic stability is equivalent to that the ``A''-matrix is Hurwitz
stable. In this subsection we study asymptotic stability of closed-loop
systems when ideal squeezers are replaced by  degenerate
parametric amplifiers (DPAs).

For simplicity, we assume that $\tilde{D}_{22}=0$ in Eq. (\ref{plant}).
As in Example \ref{Ex:squeezer}, $\tilde{S}_{u}$ is replaced by
\begin{eqnarray}
dx_{u}(t) &=&-\frac{1}{2}\left[
\begin{array}{cc}
\frac{\kappa _{u}-\epsilon _{u}}{h} & 0 \\
0 & \frac{\kappa _{u}+\epsilon _{u}}{h}%
\end{array}%
\right] x_{u}(t)dt-\sqrt{\frac{\kappa _{u}}{h}}d\tilde{u}_{o}(t) :=\tilde{A}_{u}x_{u}(t)dt+\tilde{B}_{u}d\tilde{u}_{o}(t)  \nonumber
\\
d\tilde{u}(t) &=&-\left( \sqrt{\frac{\kappa _{u}}{h}}x_{u}(t)dt+d\tilde{u}%
_{o}(t)\right) :=\tilde{C}_{u}x_{u}(t)dt+\tilde{D}_{u}d\tilde{u}_{o}(t).  \label{u}
\end{eqnarray}

Assume other ideal squeezers $\tilde{S}_{y},\tilde{S}_{vK1},\tilde{S}_{vK2}$
are replaced in a similar way. Then the closed-loop system becomes
\begin{eqnarray}
\hspace{-5mm}\left[
\begin{array}{c}
dx \\
d\xi \\
dx_{vk1} \\
dx_{vk2} \\
dx_{yo} \\
dx_{u}%
\end{array}%
\right] &\hspace{-10mm}=\hspace{-10mm}&\left[
\begin{array}{cccccc}
\tilde{A} & \tilde{B}\tilde{D}_{u}\tilde{C}_{K}+\tilde{B}_{12} & \tilde{B}%
\tilde{D}_{u}\tilde{D}_{K}\tilde{C}_{vk1} & 0 & 0 & \tilde{B}\tilde{C}_{u}
\\
\tilde{B}_{K}\tilde{D}_{y}\tilde{C}_{2}+\tilde{B}_{21} & \tilde{A}_{K} &
\tilde{B}_{K1}\tilde{C}_{vk1} & \tilde{B}_{K2}\tilde{C}_{vk2} & \tilde{B}_{K}%
\tilde{C}_{y} & 0 \\
0 & 0 & \tilde{A}_{vk1} & 0 & 0 & 0 \\
0 & 0 & 0 & \tilde{A}_{vk2} & 0 & 0 \\
\tilde{B}_{y}\tilde{C}_{2} & 0 & 0 & 0 & \tilde{A}_{y} & 0 \\
0 & \tilde{B}_{u}\tilde{C}_{K} & \tilde{B}_{u}\tilde{D}_{K}\tilde{C}_{vk1} &
0 & 0 & \tilde{A}_{u}%
\end{array}%
\right] \left[
\begin{array}{c}
x \\
\xi \\
x_{vk1} \\
x_{vk2} \\
x_{yo} \\
x_{u}%
\end{array}%
\right]  \nonumber \\
&&+\left[
\begin{array}{cccc}
\tilde{B}_{1} & \tilde{B}_{2} & \tilde{B}\tilde{D}_{u}\tilde{D}_{K}\tilde{D}%
_{vk1} & 0 \\
\tilde{B}_{K}\tilde{D}_{y}\tilde{D}_{21} & 0 & \tilde{B}_{K1}\tilde{D}_{vk1}
& \tilde{B}_{K2}\tilde{D}_{vk2} \\
0 & 0 & \tilde{B}_{vk1} & 0 \\
0 & 0 & 0 & \tilde{B}_{vk2} \\
\tilde{B}_{y}\tilde{D}_{21} & 0 & 0 & 0 \\
0 & 0 & \tilde{B}_{u}\tilde{D}_{K}\tilde{D}_{vk1} & 0%
\end{array}%
\right] \left[
\begin{array}{c}
d\tilde{B}_{in,1} \\
d\tilde{B}_{in,2} \\
d\tilde{B}_{vk1} \\
d\tilde{B}_{vk2}%
\end{array}%
\right] .  \label{cl2}
\end{eqnarray}
The following matrices are used in the sequel.
\begin{equation*}
A_{1}=\left[
\begin{array}{cc}
\tilde{A} & \tilde{B}\tilde{S}_{u}\tilde{C}_{K}+\tilde{B}_{12} \\
\tilde{B}_{K}\tilde{S}_{y}\tilde{C}_{2}+\tilde{B}_{21} & \tilde{A}_{K}%
\end{array}%
\right] ,~A_{2}=\left[
\begin{array}{cc}
\tilde{B}_{y}\tilde{C}_{2} & 0 \\
0 & \tilde{B}_{u}\tilde{C}_{K}%
\end{array}%
\right] ,
\end{equation*}%
\begin{equation*}
A_{3}=\left[
\begin{array}{cc}
0 & \tilde{B}C_{u} \\
\tilde{B}_{K}\tilde{C}_{y} & 0%
\end{array}%
\right] , ~
A_{4}=\left[
\begin{array}{cc}
\tilde{A}_{y} & 0 \\
0 & \tilde{A}_{u}%
\end{array}%
\right] ,~\Upsilon =\left[
\begin{array}{cc}
A_{1} & \frac{1}{\sqrt{h}}A_{1}A_{3}A_{4}^{-1} \\
\sqrt{h}A_{2} & A_{2}A_{3}A_{4}^{-1}%
\end{array}%
\right] .
\end{equation*}%
Notice that $\Upsilon $ is independent of $h$, while $A_{3}$ contains $1/h$.

\begin{theorem} \label{Thm:convergence}
Assume the closed-loop system (\ref{cl0}) is asymptotically stable. Then the closed-loop system (\ref{cl2}) is asymptotically stable for all $h>0$ if the matrix $\Upsilon + \Upsilon^T $
is negative definite.
\end{theorem}

\noindent \textbf{Proof.} The asymptotic stability of the closed-loop system
(\ref{cl2}) is the same as that of the following system
\begin{eqnarray}
\left[
\begin{array}{c}
dx \\
d\xi \\
dx_{yo} \\
dx_{u}%
\end{array}%
\right] &=&\left[
\begin{array}{cccc}
\tilde{A} & \tilde{B}\tilde{D}_{u}\tilde{C}_{K}+\tilde{B}_{12} & 0 & \tilde{B%
}\tilde{C}_{u} \\
\tilde{B}_{K}\tilde{D}_{y}\tilde{C}_{2}+\tilde{B}_{21} & \tilde{A}_{K} &
\tilde{B}_{K}\tilde{C}_{y} & 0 \\
\tilde{B}_{y}\tilde{C}_{2} & 0 & \tilde{A}_{y} & 0 \\
0 & \tilde{B}_{u}\tilde{C}_{K} & 0 & \tilde{A}_{u}%
\end{array}%
\right] \left[
\begin{array}{c}
x \\
\xi \\
x_{yo} \\
x_{u}%
\end{array}%
\right]  \label{system_v2} \\
&&+\left[
\begin{array}{cccc}
\tilde{B}_{1} & \tilde{B}_{2} & \tilde{B}\tilde{D}_{u}\tilde{D}_{K}\tilde{D}%
_{vk1} & 0 \\
\tilde{B}_{K}\tilde{D}_{y}\tilde{D}_{21} & 0 & \tilde{B}_{K1}\tilde{D}_{vk1}
& \tilde{B}_{K2}\tilde{D}_{vk2} \\
\tilde{B}_{y}\tilde{D}_{21} & 0 & 0 & 0 \\
0 & 0 & \tilde{B}_{u}\tilde{D}_{K}\tilde{D}_{vk1} & 0%
\end{array}%
\right] \left[
\begin{array}{c}
d\tilde{B}_{in,1} \\
d\tilde{B}_{in,2} \\
d\tilde{B}_{vk1} \\
d\tilde{B}_{vk2}%
\end{array}%
\right] .  \nonumber
\end{eqnarray}%
Therefore, in what follows we focus on system (\ref{system_v2}). Let us look
at its \textquotedblleft A\textquotedblright -matrix. Some algebraic manipulation yields
\begin{eqnarray*}
&&\left[
\begin{array}{cccc}
I & 0 & 0 & -\tilde{B}\tilde{C}_{u}A_{u}^{-1} \\
0 & I & -\tilde{B}_{K}\tilde{C}_{y}\tilde{A}_{y}^{-1} & 0 \\
0 & 0 & I & 0 \\
0 & 0 & 0 & I%
\end{array}%
\right] \left[
\begin{array}{cccc}
\tilde{A} & \tilde{B}\tilde{D}_{u}\tilde{C}_{K}+\tilde{B}_{12} & 0 & \tilde{B%
}\tilde{C}_{u} \\
\tilde{B}_{K}\tilde{D}_{y}\tilde{C}_{2}+\tilde{B}_{21} & \tilde{A}_{K} &
\tilde{B}_{K}\tilde{C}_{y} & 0 \\
\tilde{B}_{y}\tilde{C}_{2} & 0 & \tilde{A}_{y} & 0 \\
0 & \tilde{B}_{u}\tilde{C}_{K} & 0 & \tilde{A}_{u}%
\end{array}%
\right] \\
&&\times \left[
\begin{array}{cccc}
I & 0 & 0 & \tilde{B}\tilde{C}_{u}A_{u}^{-1} \\
0 & I & \tilde{B}_{K}\tilde{C}_{y}\tilde{A}_{y}^{-1} & 0 \\
0 & 0 & I & 0 \\
0 & 0 & 0 & I%
\end{array}%
\right] \\
&=&\left[
\begin{array}{cccc}
\tilde{A} & \tilde{B}\tilde{S}_{u}\tilde{C}_{K}+\tilde{B}_{12} & (\tilde{B}%
\tilde{S}_{u}\tilde{C}_{K}+\tilde{B}_{12})\tilde{B}_{K}\tilde{C}%
_{y}A_{y}^{-1} & \tilde{A}\tilde{B}\tilde{C}_{u}A_{u}^{-1} \\
\tilde{B}_{K}\tilde{S}_{y}\tilde{C}_{2}+\tilde{B}_{21} & \tilde{A}_{K} &
\tilde{A}_{K}\tilde{B}_{K}\tilde{C}_{y}A_{y}^{-1} & (\tilde{B}_{K}\tilde{S}%
_{y}\tilde{C}_{2}+\tilde{B}_{21})\tilde{B}_{1}\tilde{C}_{u}A_{u}^{-1} \\
\tilde{B}_{y}\tilde{C}_{2} & 0 & \tilde{A}_{y} & \tilde{B}_{y}\tilde{C}_{2}%
\tilde{B}\tilde{C}_{u}A_{u}^{-1} \\
0 & \tilde{B}_{u}\tilde{C}_{K} & \tilde{B}_{u}\tilde{C}_{K}\tilde{B}_{K}%
\tilde{C}_{y}A_{y}^{-1} & \tilde{A}_{u}%
\end{array}%
\right] \\
&=&\left[
\begin{array}{cc}
A_{1} & A_{1}A_{3}A_{4}^{-1} \\
A_{2} & A_{4}+A_{2}A_{3}A_{4}^{-1}%
\end{array}%
\right] .
\end{eqnarray*}%
Moreover,%
\[
\left[
\begin{array}{cc}
I & 0 \\
0 & \sqrt{h}%
\end{array}%
\right] \left[
\begin{array}{cc}
A_{1} & A_{1}A_{3}A_{4}^{-1} \\
A_{2} & A_{4}+A_{2}A_{3}A_{4}^{-1}%
\end{array}%
\right] \left[
\begin{array}{cc}
I & 0 \\
0 & \sqrt{h}%
\end{array}%
\right]^{-1} = \left[
\begin{array}{cc}
0 & 0 \\
0 & A_{4}%
\end{array}%
\right] +\Upsilon .
\]
Then by\textbf{\ Fact 5.12.3} in \cite{Ber09},
\begin{eqnarray}
\mathbf{Re}\lambda \left( \left[
\begin{array}{cc}
A_{1} & \frac{1}{\sqrt{h}}A_{1}A_{3}A_{4}^{-1} \\
\sqrt{h}A_{2} & A_{4}+A_{2}A_{3}A_{4}^{-1}%
\end{array}%
\right] \right) &\leq &\frac{1}{2}\lambda _{\max }\left( \left[
\begin{array}{cc}
0 & 0 \\
0 & 2A_{4}%
\end{array}%
\right] \right) +\frac{1}{2}\lambda _{\max }\left( \Upsilon +\Upsilon
^{T}\right)  \nonumber \\
&\leq &\frac{1}{2}\lambda _{\max }\left( \Upsilon +\Upsilon ^{T}\right). \label{eq:bernstein}
\end{eqnarray}%
Note the right-hand side of (\ref{eq:bernstein}) is independent of $h$.
Consequently, if the matrix $\Upsilon +\Upsilon ^{T}$ is negative definite,
then the closed-loop system (\ref{cl2}) is asymptotically stable for all $h >0$, so is the
closed-loop system (\ref{cl0}). $\Box $


\subsection{Performance convergence} \label{sec:perf_convg}

\begin{figure}[tbh]
\epsfxsize=2.0in
\par
\epsfclipon
\par
\centerline{\epsffile{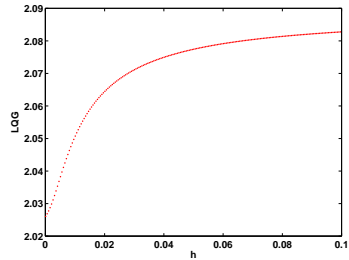}}
\caption{Closed-loop quantum LQG Performance with replacement of ideal squeezers by degenerate parametric amplifiers}
\label{h2}
\end{figure}

Pertaining to the study in Sec.~\ref{sec:limit}, if we replace the squeezers
$\tilde{S}_{u}$, $\tilde{S}_{vK1}$, $\tilde{S}_{vK2}$, and $\tilde{S}_{y}$
by DPAs respectively, then the closed-loop system (\ref{cl2}) is obtained. According to Theorem \ref{Thm:convergence}, simple calculations show that this closed-loop system is asymptotically
stable for all $h>0$. Fig.~\ref{h2} shows how the closed-loop LQG
performance changes as a function of $h$. It can be seen that, as $h$ goes
to zero, the performance converges to $2.02588$, which is slightly bigger
than $2.0004$, but still better than $2.03013$, the vacuum case given by Theorem \ref{thm:vacuum}.

\section{Conclusion}

\label{sec:conclusion}

Recent years have seen a considerable amount of work on quantum networks
theory, e.g., system interconnection \cite{YK03,GJ08,GGY08,GJ09,GJN10}, $%
H^{\infty}$ control \cite{JNP08,ZJ11}, LQG control \cite{NJP09,ZJ11}, synthesis theory \cite%
{NJD09,Nur10}, dissipation theory and direct coupling \cite{JG10,ZJ11}, laboratory demonstrations \cite{Mabuchi08}, among others. This
paper fits into such general picture by presenting how to use phase shift and squeezing in control design for linear quantum optical
feedback networks. In addition, the controller parametrization and
two-stage optimization procedure may be useful in the optimal design of
quantum optical networks, their advantage for engineering non-classical
correlation has been illustrated in the study of coherent LQG quantum
feedback control.


\section*{Acknowledgment}

The first author wishes to thank H.I Nurdin for his helpful discussions on
quantum LQG control and comments on the paper.

%

\end{document}